# EFFECT OF COPPER CONTENT ON THERMAL AND MECHANICAL PROPERTIES OF EUTECTOID Zn-Al ALLOY

Md. Arifur Rahman Khan[a,*], Toufiq Ahmed[a], Md. Khairul Islam[a], Sajib Aninda Dhar[b], M. R. Qadir[b]

[a]*Institute of Mining, Mineralogy and Metallurgy, BCSIR, Joypurhat-5900, Bangladesh*
[b]*Pilot Plant and Process Development Centre, BCSIR, Dhaka-1205, Bangladesh*

**Abstract**

Zn-Al alloys have become one of the major engineering alloys among commercially available alloys. This study was conducted on eutectoid composition of Zn-Al alloy with an observation of the effect of copper addition. For this purpose, one eutectoid (Zn-22wt%Al) and three other alloys, adding 1wt%, 3wt% and 5wt% copper with this eutectoid composition, were molded in permanent metal mould. Microscopic studies exhibited varied grains which confirmed the formation of different phases. Moreover, the formation of different phases in micro study was supported by XRD analysis. Hardness of the samples were tested on Rockwell B scale and it was observed that the hardness of these alloys was substantially increased with the addition of copper. With increasing amount of copper, phase changing temperature of the alloys reveals a growing trend, which was observed by DTA analysis. From this study it was concluded that addition of copper can significantly add to the mechanical properties of Zn-Al alloys.

## 1. INTRODUCTION

Zinc-aluminium binary alloys have been studied for many years as a replacement of costly alloys. ZAMAK, ALZEN and ZA alloys are the most important commercial alloys [1-4]. Getting optimum property from Zinc-aluminium binary alloys is a constraint for engineering applications. Zn has been used for galvanization for many years. Considering the requirement of different properties of coating material for varied metals and alloys, development of other coating materials is a necessity. Zn-Al eutectoid alloys provide optimum properties for many engineering applications. This eutectoid alloy has been extensively used for galvanization [5]. Besides the compatibility of coating material with coated material, some other properties like strength, hardness, wear rate and friction coefficient etc. are also considered as a prime requirement for developing coating material. Moreover, low melting point of coating material will reduce energy consumption during melting which is a prime factor for commercial applications.

Various alloying elements were added to improve the properties of zinc aluminium binary alloys. Cu is one of the most effective elements towards improving mechanical and tribological properties [6-12]. Addition of Cu to Zn-Al alloys shows improved properties along with low melting point compared to other commercial alloys. Cu addition improves castability, superplasticity, strength and hardness, fatigue strength, wear rate, friction coefficient and manufacturing costs [13-16]. The purpose of this study is to investigate the effect of Cu on eutectoid microstructure, thermal and mechanical properties.

## 2. EXPERIMENTAL PROCEDURE

One eutectoid alloy and three Cu added eutectoid alloys were produced from commercially pure aluminium, zinc and pure copper. Sample 1, eutectoid alloy is Zn-22Al. Other three alloys, sample 2, sample 3, sample 4 were produced by addition of respectively 1wt%, 3wt% and 5wt% Cu to this eutectoid composition alloy. Alloys were melted in a graphite crucible using a muffle furnace and poured at around $600^0$C temperature in a permanent mould. Samples of all kinds were produced by sectioning and grinding using emery paper and finally polished using gamma alumina powder for optical microscopic study. Samples of optical microscopic study were used for XRD analysis using BRUKER D8 advance (Germany). Phase identification was confirmed by overlapping reference data on the XRD result. DTA was done by SII EXSTAR TG/DTA 6300 where powder samples of all types were used. Hardness was measured in Rockwell B scale using hardness tester of SHIMADZU. For hardness measurement, polished samples of microscopic study were used. Throughout the experimental procedure samples were preserved in desiccator to save from moisture contamination and surface corrosion.

### 3. RESULTS AND DISCUSSIONS

### 3.1 MICROSTRUCTURE STUDY

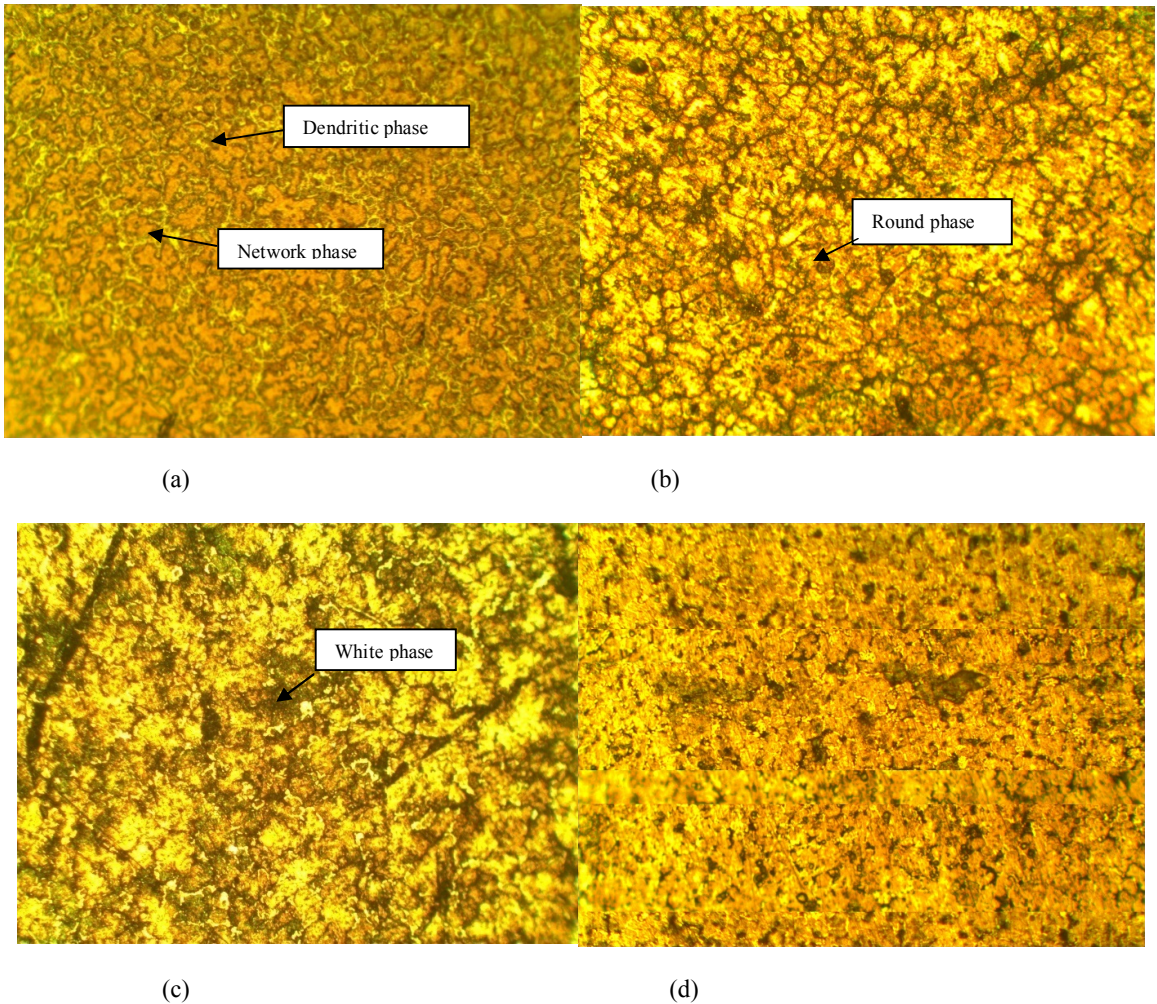

Fig 1: Microstructure (200x) of (a) Sample 1 (b) Sample 2 (c) Sample 3 (d) Sample 4

Fig 1 shows the microstructural image of alloyed and unalloyed eutectoid composition. For sample 1, fig 1(a) reveals two different phases, dendritic phase of dark color and light colored network structure phase. As alloying element Cu is added to sample 2, 3, 4, absence of this network structure is noticed. For sample 2, a new phase of nearly round shape is seen to be appeared (fig b). For sample 3 where 3wt% Cu was added, a new phase of white color is observed (fig c). As per the XRD results this new phase may be of $Cu_3Al_2$. This phase is absent in fig b, may be due to the low percentage addition (1wt%) of Cu. Microstructure of sample 4 shows more phases of white color along with nearly round shape phase of fig b. This variation was studied by XRD which shows existence of Cu rich $Cu_{0.8}Zn_{0.2}$ phase along with $Cu_3Al_2$ phase of sample 3.

**3.2 XRD ANALYSIS**

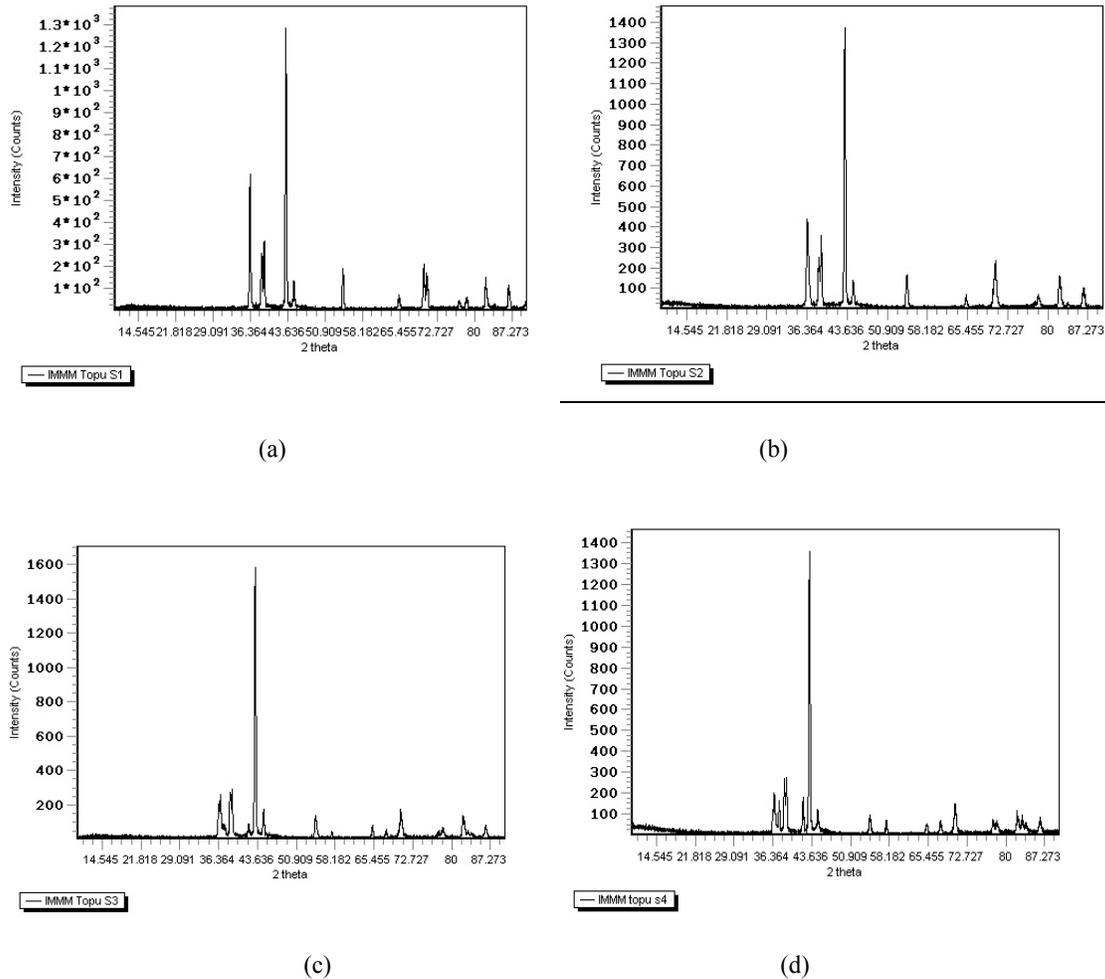

(a)　　　　　　　　　　　　　　　　(b)

(c)　　　　　　　　　　　　　　　　(d)

Fig 2: XRD results of (a) Sample 1 (b) Sample 2 (c) Sample 3 (d) Sample 4

XRD result showed the presence of various compound formed by Zn, Al and Cu. Eutectoid Sample 1 showed presence of Zn and $(Al_{19}Zn)_{0.2}$ phase. Sample 2 reflects the same phases of sample 1 where 1%wt Cu was added. But with more percentage of Cu addition new phases are found.

XRD data of sample 3 confirmed the presence of compound $(Al_{19}Zn)_{0.2}$, Zn, $Cu_3Al_2$. Moreover, data of sample 4 showed the presence of an additional compound $Cu_{0.8}Zn_{0.2}$ with all the compounds of sample 3. Sample 3 contains 3%wt of Cu while sample 4 has 5%wt of Cu. From this observation we may say that Cu first forms alloy with Al then with abundance of Cu we will get copper zinc phase.

Crystallite size was calculated from XRD data considering the first peak of each figure by following scherrer equation:

$$\tau = \frac{k\lambda}{\beta \cos\theta}$$

Here,
$\tau$ = mean size of the ordered (crystalline) domains,

K= dimensionless shape factor, with a value close to unity. The shape factor has a typical value of about 0.9
λ = the X-ray wavelength;
β = the line broadening at half the maximum intensity (FWHM) (in radians), denoted as
θ = the Bragg angle (in degrees).

Table 1: Data of β and θ for scherrer equation

| Sample Identity | β (in degree) | 2θ (in degree) |
|---|---|---|
| Sample 1 | 0.2808 | 36.2309 |
| Sample 2 | 0.4610 | 36.4639 |
| Sample 3 | 0.6783 | 36.6598 |
| Sample 4 | 0.5460 | 36.6603 |

Calculation of crystallite size using table 1 for all types of samples provides interesting result. Crystallite sizes are observed to be decreased as addition of Cu increases. According to scherrer equation crystallite sizes are 297.67 $A^0$, 181.42 $A^0$, 123.38 $A^0$ and 153.28 $A^0$ for sample 1, 2, 3 and 4 respectively. There is a variation in decreasing trend of crystallite size for sample 4 which is showing higher size than sample 3. As per the Hall Patch relationship it is known that strength is inversely proportional to the grain size. From scherrer calculation strength may increase up to 3wt% Cu addition. If Cu is added beyond this level, as per the calculation sample 4 will show low strength. This is supported by previous study of some other researcher [17-18].

### 3.3 DTA ANALYSIS

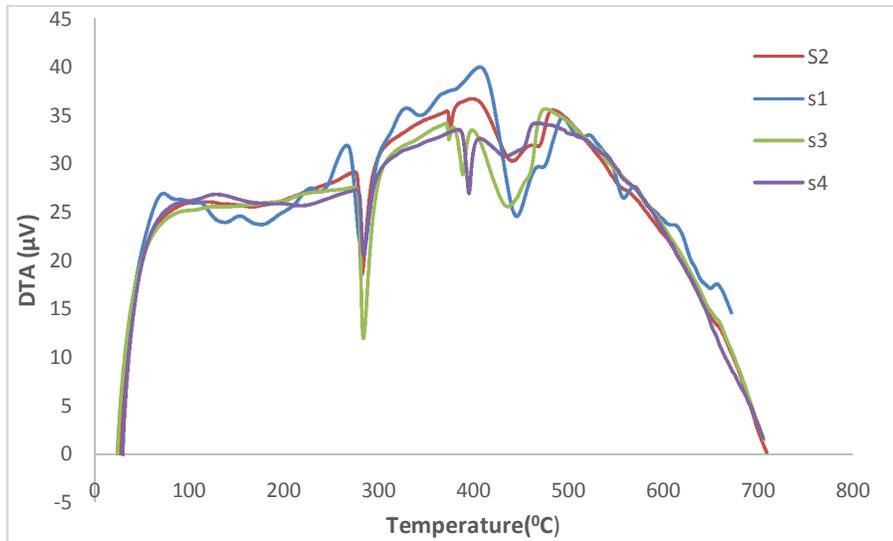

Fig 3: DTA (μv) VS TEMPERATURE

Fig 3 shows variation of phase changing temperatures for all the samples. For the sample 1, eutectoid composition, DTA result should provide a peak at temperature $275^0C$ for eutectoid phase transformation. In this experiment eutectoid temperature was found at $281.6^0C$ (fig a), the variation of temperature may be the effect of using commercial grade metals. Other results from fig b,c and d, show shifting of eutectoid temperature to higher temperature as content of Cu increases. But the effect of Cu on eutectoid temperature is not very prominent. On the other hand, all the samples show peaks at around $440^0C$. This may be explained by the transformation temperature of eutectoid phase to Al rich phase and liquid phase. For unalloyed eutectoid composition, sample 1, this temperature was found at $444.6^0C$ but as the alloying element Cu increases this temperature shows a decreasing trend. This eutectoid to Al rich phase transformation temperature was found around $440^0C$, $436^0C$ and $433^0C$ for samples respectively 2, 3 and 4.

### 3.4 HARDNESS

Rockwell hardness test (Scale B) was done for all the samples with applied load 100Kg.

Table 2: Average Rockwell hardness data and standard deviation

| Sample Identity | HRB (Average) | Standard Deviation |
|---|---|---|
| Sample 1 | 30.5 | 0.86 |
| Sample 2 | 53.25 | 1.08 |
| Sample 3 | 54.75 | 0.82 |
| Sample 4 | 61.75 | 2.16 |

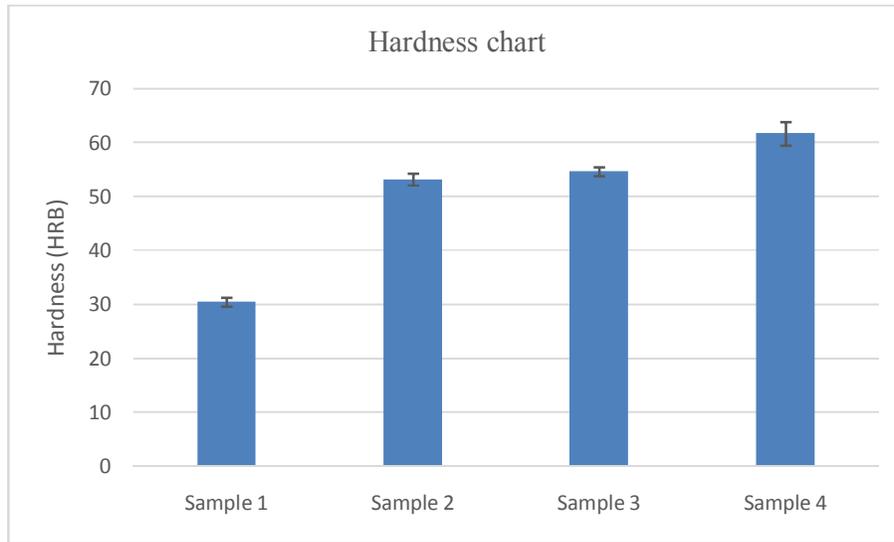

Fig 4: Comparison of hardness of all samples.

From Fig 4, we see that there is a sequential increment of hardness with addition of Cu. Moreover, there is a remarkable change in sample 2 compared with sample 1. Sample 1 was of pure eutectoid composition while sample 2, 3 and 4 was of eutectoid composition with addition of Cu respectively 1%wt, 3%wt, and 5%wt. This may be the result of crystal structure. Al and Cu both are FCC crystal while Zn is HCP crystal. XRD result has confirmed the presence of $(Al_{19}Zn)_{0.2}$, $Cu_3Al_2$ phases where first one is FCC crystal and the last one is HCP crystal. Compared with sample 1, sample 3 contains Al which is now in HCP structure. Hexagonal crystal structure acts as load bearing phase providing hardening effect and thus offers better wear resistance [7,9 and 19]. From metallurgical point of view, FCC crystal has 12 slip systems while HCP crystal has only 3 slip systems. This limits the plastic deformation of HCP structure compared with FCC structure which makes the alloy (sample 3) harder than sample 1 and sample 2. On the other hand, with all the phases present in sample 3, sample 4 has one more phase, $Cu_{0.8}Zn_{0.2}$, which is also a HCP structure. So hardness increases for the same reason mentioned above. Increment of hardness from sample 3 to sample 4 is not very prominent. This point may be explained by the presence of only one more HCP phase, $Cu_{0.8}Zn_{0.2}$. Sample 3 and sample 4 have similar phases except the additional $Cu_{0.8}Zn_{0.2}$ of sample 4. Since Cu is added in limited amount, the amount of this phase will not be very high. Moreover, $Cu_{0.8}Zn_{0.2}$ is formed after the formation of $Cu_3Al_2$. So increment of hardness is limited. Furthermore, formation of these Cu rich phases at grain boundaries will make dislocation movement difficult providing pinning effect. Which are also congruent with hardness increment property. Wear loss is inversely proportional to strength and hardness [20 and 21]. So better wear resistance is expected from these alloys.

## 4. CONCLUSION

1. Addition of Cu to Zn-Al eutectoid alloy changes system from network structure phase to Cu rich new phases which ware confirmed by XRD results.
2. XRD results show presence of more hexagonal phases as Cu increases which make alloys harder than binary alloy. Moreover Crystallite sizes show a decreasing trend for Cu addition up to 3%, beyond this crystallite size increases.
3. Cu has little effect on eutectoid temperature shifting but transformation temperature of eutectoid to Al rich phase and liquid phase shows a decreasing trend as addition of Cu increases.
4. Hardness of the alloys increases with the increase of Cu addition for all the samples which will provide better wear resistance than binary eutectoid alloy.